\documentstyle[12pt]{article}

\newcommand{\be}{\begin{equation}}
\newcommand{\ee}{\end{equation}}
\newcount\sectionnumber
\def\sect
{\global\equationnumber=0
\global\advance\sectionnumber by 1
\the\sectionnumber . }
\newcount\equationnumber
\def \num
{\eqno{\global\advance\equationnumber by 1
\left(\the\sectionnumber .\the\equationnumber \right)}}

\def\wave{ \vcenter{\vbox{\hrule height.3pt \hbox{\vrule width.3pt height9pt
\kern9pt \vrule width.3pt} \hrule height.3pt}} \, }
\def\s{ \hspace{1mm} }
\def\b{ \hspace{2mm} }
\def\B{ \hspace{4mm} }
\def\EXP{ {\rm e} }
\def\comma{ \b , }
\def\period{ \b . }
\def\Del{ \nabla }
\def\di{ \partial }
\def\de{ {\rm d} }
\def\half{ \frac{1}{2} }

\def\const{ 8\,\pi\,G }
\def\minus{ \mbox - \, }
\def\invs{ ^{\mbox - \,1} }
\def\Dt{ \frac{d}{dt} }

\def\Dx{ \frac{d}{dx} }
\def\dit{ \partial_t }
\def\dix{ \partial_x }
\def\ditt{ \partial_{tt} }
\def\ditx{ \partial_{tx} }
\def\dixx{ \partial_{xx} }
\def\Ta{ \tilde{T} }
\def\Tb{ \check{T} }
\def\Ja{ \tilde{J} }
\def\Jb{ \check{J} }
\def\PhiO{ \stackrel{\scriptscriptstyle 0}{\Phi} }
\def\PhiA{ \stackrel{\scriptscriptstyle 1}{\Phi} }
\def\PhiB{ \stackrel{\scriptscriptstyle 2}{\Phi} }
\def\PhiC{ \stackrel{\scriptscriptstyle 3}{\Phi} }
\def\PhiD{ \stackrel{\scriptscriptstyle 4}{\Phi} }
\def\gttB{ \stackrel{\scriptscriptstyle 2}{g}_{00} }
\def\gtxC{ \stackrel{\scriptscriptstyle 3}{g}_{10} }
\def\gttD{ \stackrel{\scriptscriptstyle 4}{g}_{00} }
\def\cJO{ \stackrel{\scriptscriptstyle 0}{\cal J} }
\def\cJA{ \stackrel{\scriptscriptstyle 1}{\cal J} }
\def\cJB{ \stackrel{\scriptscriptstyle 2}{\cal J} }

\def\TO{ \stackrel{\scriptscriptstyle 0}{\cal T} }
\def\TA{ \stackrel{\scriptscriptstyle 1}{\cal T} }
\def\TB{ \stackrel{\scriptscriptstyle 2}{\cal T} }

\def\JO{ \stackrel{\scriptscriptstyle 0}{J} }
\def\JB{ \stackrel{\scriptscriptstyle 2}{J} }
\def\JD{ \stackrel{\scriptscriptstyle 4}{J} }
\def\TttO{ \stackrel{\scriptscriptstyle 0}{T}_{00} }
\def\TttB{ \stackrel{\scriptscriptstyle 2}{T}_{00} }
\def\TtxA{ \stackrel{\scriptscriptstyle 1}{T}_{10} }
\def\TxxB{ \stackrel{\scriptscriptstyle 2}{T}_{11} }

\def\cJ{ \cal J }
\def\gH{ \hat{g} }
\def\hH{ \hat{h} }
\def\phiH{ \phi }
\def\vphiH{ \varphi }
\def\DelH{ \hat{\nabla} }

\begin{document}
\pagenumbering{roman}
\title{Gravitation and Cosmology in Generalized (1+1)-dimensional dilaton
gravity}

\author{J. S. F. Chan \\
Department of Applied Mathematics, \\
University of Waterloo, \\
Waterloo, Ontario, Canada \\
\vspace{10pt}
N2L 3G1 \\
R. B. Mann\\
Department of Physics, \\
University of Waterloo, \\
Waterloo, Ontario, Canada \\
N2L 3G1}

\date{}
\maketitle

\begin{abstract}
  The actions of the ``$R=T$'' and string-inspired theories of
  gravity in (1+1) dimensions are generalized into one single action
  which is characterized by two functions.  We discuss differing
interpretations of
  the matter stress-energy tensor, and show how two such different
interpretations
  can yield  two different
  sets of field equations from this action. The
  weak-field approximation, post-Newtonian expansion, hydrostatic
  equilibrium state of star and two-dimensional cosmology are
  studied separately by using the two sets of field equations. Some
  properties in the ``$R=T$'' and string-inspired theories are shown
  to be generic in the theory induced by the generalized action.
\end{abstract}

\pagenumbering{arabic}

\section{Introduction} \label{Intro}
  \bigskip

  In (1+1)-dimensional spacetime, the metric tensor $g_{\mu \nu}$ has
  only three different components, two of which may be eliminated by
  a choice of coordinates. This fact reduces the complexity of
  metric-related computations substantially. Moreover, the field
  equations are expected to be easier to solve than those in ordinary
  (3+1)-dimensional spacetime because all fields depend on at most
  two variables.

  As a result, theorists have found two-dimensional gravity an
  attractive theoretical laboratory for gaining insight into issues
  in semi-classical and quantum gravity \cite{Brown}, \cite{Jackiw}.
  Despite its relatively easy computability, the  two-dimensional
  setting suffers a serious drawback: the Einstein field equations
  are no longer a feasible model of the spacetime because the
  Einstein tensor is identically  zero, yielding a theory without
  any dynamical content. In order to cope with this difficulty,
  several theories of gravity in two dimensions have been proposed
  by relativists. Recently, two such theories, referred to as
  ``$R = T$'' theory \cite{Mann} and string-inspired theory
  \cite{Callan}, have attracted attention for a variety of reasons,
  including the fact that they have interesting classical limits
  \cite{Mann1,Ross} and admit black hole solutions.

  The ``$R = T$'' theory can be derived from the action
  \begin{eqnarray}
    S & = &
    \int \de^2x\,\sqrt{\minus g}\,\left[\,
    \half\,g^{\mu \nu}\,\Del_\mu \psi\,\Del_\nu \psi
    + \psi\,R + {\cal L}_M\,\right] \comma \label{S1E1}
  \end{eqnarray}
  where the scalar field $\psi$ is an auxiliary field, so called
  because the classical field equations may be rewritten in a form
  that removes the $\psi$-dependence from the equations governing
  the evolution of the metric and matter fields. These are
  \begin{eqnarray*}
    R & = & \const\,g^{\mu \nu}\,T_{\mu \nu}
  \end{eqnarray*}
  and the equation for the covariant conservation of the
  stress-energy tensor.

  The string-inspired theory arises from a noncritical string theory
  in (1+1) dimensions. By setting the one-loop $\beta$ function of
  the bosonic $\sigma$ model with two target spacetime dimensions
  to zero, the effective target space action becomes
  \begin{eqnarray}
    S & = &
    \int \de^2x\,\sqrt{\minus g}\,
    \EXP^{\minus 2\,\phi}\,\left\{\,\left[\,4\,\Del_\lambda \phi\,
    \Del^\lambda \phi
    + R + J\,\right] + {\cal L}_M\,\right\} \comma \label{S1E2}
  \end{eqnarray}
  where $\phi$ is the dilaton field and the field equations of this
  action are
  \begin{eqnarray*}
    J & = & \minus R - 4\,\Del^2 \phi + 4\,(\Del \phi)^2 \comma \\
    8\,\pi\,G\,T_{\mu \nu} & = &
    \EXP^{\minus 2\,\phi}\,\left(\,R_{\mu \nu} + 2\,\Del_\mu \Del_\nu
\phi\,\right) \period
  \end{eqnarray*}

  The actions (\ref{S1E1}) and (\ref{S1E2}) are particular types of
  dilaton theories of gravity, theories in which a scalar field (the
  dilaton) couples non-minimally to gravity via the Ricci scalar.
  These theories have a number of interesting properties, both in
  terms of their classical conservation laws \cite{Mann2} and their
  quantum structure \cite{Banks}. However (except in a few special
  cases \cite{Mann1,Ross}) their detailed classical properties have
  never been systematically analyzed. In this paper we undertake
  this task.

  A wide class of dilation gravity theories can be obtained if we
  generalize the actions for the ``$R=T$'' and string-inspired
  theories into the action
  \begin{eqnarray}
    S & = & S_g - S_s \comma \label{S1E3}
  \end{eqnarray}
  where
  \begin{eqnarray*}
    S_g :=
    \int \de^2x\,\sqrt{\minus g}\,\left[\,H(\Phi)\,(\Del \Phi)^2
    + D(\Phi)\,R\,\right] & {\rm and} &
    S_s := \int \de^2x\,\sqrt{\minus g}\,V
  \end{eqnarray*}
  because the functions $H(\Phi)$ and $D(\Phi)$ can be used to
  characterize the theory \cite{Mann2}. For example, in ``$R=T$''
  theory, $H(\Phi) = 1/2$ and $D(\Phi) = \Phi$, and in the
  string-inspired theory, $H(\Phi) = 4\,D(\Phi) = 4\,\exp(\minus 2\,\Phi)$.
  The term $V$, which is interpreted as the potential density for
  the matter and dilaton sources, may depend on $\Phi$, derivatives
  of $\Phi$, the metric tensor and other matter field. The variations
  of (\ref{S1E3}) with respect to the contravariant metric
  $g^{\mu \nu}$ and the dilaton field  $\Phi$ give
  \be
    \frac{\delta S_s}{\sqrt{\minus g}\,\delta g^{\mu \nu}} =
    H(\Phi)\,\left[ \Del_\mu \Phi\,\Del_\nu \Phi - \half\,(\Del \Phi)^2\,g_{\mu
\nu} \right]
    - \Del_\mu \Del_\nu D(\Phi)
    + \Del^2 D(\Phi)\,g_{\mu \nu} \comma \label{S1E4}
  \ee
  \be
    \frac{\delta S_s}{\sqrt{\minus g}\,\delta \Phi} =
    D'(\Phi)\,R - 2\,H(\Phi)\,\Del^2 \Phi
    - H'(\Phi)\,(\Del \Phi)^2 \period \label{S1E5}
  \ee
  Since the Ricci scalar is the only measure of curvature in $(1+1)$
  dimensions, we will require it to appear in at least one field
  equation. Consequently we assume the following throughout the paper:
  \begin{eqnarray}
    D'(\Phi) \neq 0 && \forall \quad \Phi \label{AS1}
  \end{eqnarray}
  but otherwise consider $D(\Phi)$ to be an arbitrary function of $\Phi$.

  Our analysis of the classical properties of dilaton gravity will
  therefore be based upon the action (\ref{S1E3}), with $H$ and $D$
  arbitrary functions of the dilaton field $\Phi$ (modulo (\ref{AS1}))
  and $V$ an arbitrary function of the dilaton field and any other
  matter fields $\Psi$ in the system.  This is the  most general
  action linear in the curvature and quadratic in the derivatives of
  $\Phi$ and the matter fields. The action (\ref{S1E3}) actually only
  depends upon the function $V(\Phi;\Psi)$ since reparametrizations of
  $\Phi$ accompanied by $\Phi$-dependent Weyl rescalings of the metric
  allow one to relate models with different $H$'s and $D$'s \cite{Mann2}.
  Only the overall sign and critical points of $H$ and $D$ contain
  reparametrization invariant information \cite{Banks}. The matter
  potential $V$ breaks Weyl invariance, and so the field equations
  (\ref{S1E4}) and (\ref{S1E5}) determine the evolution of the spacetime
  metric and matter fields. General coordinate invariance implies that
  locally the evolution of the metric is determined by the evolution of
  its conformal factor.

  This arbitrariness in the potential $V$ yields some ambiguity in
  terms of how the left-hand-sides of (\ref{S1E4}) and (\ref{S1E5})
  are interpreted. For example, if the variations of (\ref{S1E4})
  and (\ref{S1E5}) equal
  \begin{eqnarray*}
    \frac{\delta S_s}{\sqrt{\minus g}\,\delta g^{\mu \nu}} \b := \b
    \const\,\Ta_{\mu \nu} \B & \B {\rm and} \B & \B
    \frac{\delta S_s}{\sqrt{\minus g}\,\delta \Phi} \b := \b \Ja \comma
  \end{eqnarray*}
  we obtain the field equations
  \be
    \Ja =
    D'(\Phi)\,R - 2\,H(\Phi)\,\Del^2 \Phi
    - H'(\Phi)\,(\Del \Phi)^2 \comma \label{SFE1}
  \ee
  \be
    \const\,\Ta_{\mu \nu} =
    H(\Phi)\,\left[\,\Del_\mu \Phi\,\Del_\nu \Phi - \half\,(\Del
\Phi)^2\,g_{\mu \nu}\,\right]
    - \Del_\mu \Del_\nu D(\Phi)
    + \Del^2 D(\Phi)\,g_{\mu \nu} \period \label{TFE1}
  \ee
  We shall refer to the system (\ref{SFE1}) and (\ref{TFE1}) as the
  {\it type-I} field equations, and interpret $\Ta_{\mu \nu}$ and
  $\Ja$ as the stress-energy-momentum tensor of the matter field
  and the source of dilaton respectively. The divergence of
  (\ref{TFE1}) reads
  \begin{eqnarray}
    \const\,\Del^\nu \Ta_{\mu \nu} & = &
    \minus \half\,\Ja\,\Del_\mu \Phi \period \label{CL1}
  \end{eqnarray}
  Note that the dilaton source $\Ja$ vanishes if the tensor
  $\Ta_{\mu \nu}$ is covariantly conserved.

  Alternatively, consider the restriction
  \begin{eqnarray}
    H(\Phi) & = & k\,D'(\Phi) \comma \label{S1E6}
  \end{eqnarray}
  where $k$ is a non-zero constant. In this case equations
  (\ref{S1E4}) and (\ref{S1E5}) can be rearranged as
  $$
    \frac{\delta S_s}{\sqrt{\minus g}\,\delta \Phi} \b = \b
    D'(\Phi)\,R - 2\,k\,D'(\Phi)\,\Del^2 \Phi
    - k\,D''(\Phi)\,(\Del \Phi)^2 \comma
  $$
  \begin{eqnarray}
    && \frac{\delta S_s}{\sqrt{\minus g}\,\delta g^{\mu \nu}}
    + \frac{g_{\mu \nu}}{2\,k}\,\frac{\delta S_s}{\sqrt{\minus g}\,\delta \Phi}
\b = \b
    \frac{1}{k}\,D'(\Phi)\,R_{\mu \nu}
    - D'(\Phi)\,\Del_\mu \Del_\nu \Phi \nonumber \\ &&
    \hspace{20mm} + \s \left[\,k\,D'(\Phi) -
D''(\Phi)\,\right]\,\left[\,\Del_\mu \Phi\,\Del_\nu \Phi - \half\,(\Del
\Phi)^2\,g_{\mu \nu}\,\right] \period \B \B \label{S1E7}
  \end{eqnarray}
  If the left side of (\ref{S1E7}) is defined to be
  $\const\,\Tb_{\mu \nu}$, the equation will yield
  \begin{eqnarray*}
    \const\,\Del^\nu \Tb_{\mu \nu} & = &
    \frac{1}{k}\,\Del_\mu \left[\,\half\,\frac{\delta S_s}{\sqrt{\minus
g}\,\delta \Phi}\,\EXP^{\minus k\,\Phi}\,\right]\,\EXP^{k\,\Phi} \period
  \end{eqnarray*}
  This result suggests that the term inside the square brackets can
  be identified as the dilaton source $\Jb$ such that the divergence
  of the stress-energy-momentum tensor becomes simply
  \begin{eqnarray}
    \const\,\Del^\nu \Tb_{\mu \nu} & = &
    \frac{1}{k}\,\EXP^{k\,\Phi}\,\Del_\mu \Jb \period \label{CL2}
  \end{eqnarray}
  Equation (\ref{CL2}) implies that if the stress-energy-momentum
  tensor $\Tb_{\mu \nu}$ obeys the local conservation laws, the
  dilaton source $\Jb$ must be a constant which is not necessarily
  zero, in contrast to the previous case. Consequently  this interpretaton,
along with the
  restriction (\ref{S1E6}),  yields another set of field equations as
  the following:
  \begin{eqnarray}
    \Jb & = &
    \half\,\EXP^{\minus k\,\Phi}\,\left[\,D'(\Phi)\,R - 2\,k\,D'(\Phi)\,\Del^2
\Phi
    - k\,D''(\Phi)\,(\Del \Phi)^2\,\right] \comma \B \label{SFE2} \\
    \const\,\Tb_{\mu \nu} & = &
    \frac{1}{k}\,D'(\Phi)\,R_{\mu \nu}
    - D'(\Phi)\,\Del_\mu \Del_\nu \Phi \nonumber \\ &&
    + \s \left[\,k\,D'(\Phi)
    - D''(\Phi)\,\right]\,\left[\,\Del_\mu \Phi\,\Del_\nu \Phi
    - \half\,(\Del \Phi)^2\,g_{\mu \nu}\,\right] \period \B \B \label{TFE2}
  \end{eqnarray}

  These new equations will henceforth be denoted as {\it type-II}
  field equations. As the proportionality constant $k$ in condition
  (\ref{S1E6}) is non-zero, the type-II field equations can be used
  only when $H \neq 0$. If $H$ is zero, only type-I field equations
  are available. It is worthwhile pointing out that the type-II
  field equations include not only the ``$R = T$'' theory but also
  the string-inspired theory as well.  The ambiguity giving rise to
  the two types of field equations originates from lack of knowledge
  of the potential density $V$. Clearly many other interpretations
  of the metric variation of the matter action $S_s$ in terms of
  the stress-energy and dilation currents are possible. However, in
  the absence of additional information on $D$, $H$ and $V$, the
  type-I and type-II equations cover a large class of matter
  couplings (including both the $R=T$ and string-theoretic cases)
  and we shall henceforth consider only these two choices.

  The outline of our paper is as follows. We first consider the
  weak-field approximations and post-Newtonian expansions of the
  field equations in sections~\ref{WFA} and \ref{PNA} separately.
  In section~\ref{Stellar}, the interior structure of a `star' in
  (1+1)-dimensional spacetime will be investigated. Cosmological
  solutions derived from the field equations will be considered in
  section~\ref{Cosmo} for dust-filled, radiation-filled and
  inflationary universes.  Finally some conclusions will be drawn
  in section~\ref{Conclus}.

  A given choice of the characteristic functions  $D$, $H$, and $V$
  corresponds to a given theory. All computations are carried out
  without imposing any restrictions upon these functions except for
  the constraint (\ref{AS1}) (which guarantees that the metric has
  non-trivial evolution equations). Of course the freedom of
  choosing the characteristic functions permits the formulation of
  inverse problems. For example, what choice of characteristic
  functions is required for a desired spacetime evolution? An
  example of inverse problem will be shown in section~\ref{Cosmo.3}.


\section{Weak Field Approximation} \label{WFA}
  \bigskip

  We consider here a weak-field expansion of the metric about some fixed
  background for the type-I and type-II systems separately.

  \subsection{Type I Equations} \label{WFA.1}
    Consider a weak-field expansion about some background metric.
    Under a conformal transformation
    \begin{eqnarray*}
      g_{\mu \nu} & = &
      \exp\,\left(\,\minus \int^{\Phi} \frac{H(u)}{D'(u)}\,\de
u\,\right)\,\gH_{\mu \nu} \comma
    \end{eqnarray*}
    the field equations (\ref{SFE1}) and (\ref{TFE1}) can be
    transformed into
    \be
      \Ja =
      \exp\!\left(\,\int^{\Phi} \frac{H(u)}{D'(u)}\,\de
u\,\right)\,\left[\,D'(\Phi)\,\hat{R} - \frac{H(\Phi)}{D'(\Phi)}\,\DelH^2
D(\Phi)\,\right] \comma \B \label{S2E1}
    \ee
    \be
      \const\,\Ta_{\mu \nu} =
      \DelH^2 D(\Phi)\,\gH_{\mu \nu}
      - \DelH_\mu\!\DelH_\nu D(\Phi) \comma \label{S2E2}
    \ee
    where every term with the hat is interpreted with respect to the
    conformal metric $\gH_{\mu \nu}$. In carrying out the weak field
    calculation, it will be easier to use the transformed field
    equations (\ref{S2E1}) and (\ref{S2E2}) than to use the original
    equations (\ref{SFE1}) and (\ref{TFE1}) because the $H$-dependence
    in (\ref{TFE1}) is transformed away. This property of
    $H$-independence in (\ref{S2E2}) renders the vacuum calculation
    tractable without knowing any properties of the function $D$.

    Now we let the dilaton source $\Ja$ be
    \begin{eqnarray}
      \Ja & = & \epsilon\,c + \cJ \comma \label{S2E3}
    \end{eqnarray}
    where $c$ is a non-zero constant and $\epsilon$ is a parameter
    which may either be 0 or 1, depending on the choice of dilaton
    vacuum. As the term $\cJ$ represents the non-vacuum part of $\Ja$,
    equation (\ref{S2E3}) indeed splits the dilaton source into
    ``vacuum'' and ``non-vacuum'' parts. Similarly, we expand the
    fields $\gH_{\mu \nu}$ and $\Phi$ via
    \begin{eqnarray*}
      \gH_{\mu \nu} \b = \b \eta_{\mu \nu} + \hH_{\mu \nu} & \B {\rm and} \B &
      \Phi = \phiH + \vphiH \comma
    \end{eqnarray*}
    where $\hH_{\mu \nu}$ and $\vphiH$ are perturbations of the fields
    from the vacuum fields $\eta_{\mu \nu}$ and $\phiH$. As is
    standard in perturbation theory, we take $|\hH_{\mu \nu}| << 1$
    and $|\vphiH|<<1$.

    Since the flat metric $\eta_{\mu \nu}$ and $\phiH$ are supposed
    to be the solutions in vacuum, they satisfy the equations
    \be
      \epsilon\,c =
      \minus \frac{H(\phiH)}{D'(\phiH)}\,\wave
D(\phiH)\,\exp\!\left(\,\int^{\phiH} \frac{H(u)}{D'(u)}\,\de u\,\right) \comma
\label{S2E4}
    \ee
    \be
      \wave D(\phiH)\,\eta_{\mu \nu} =
      \di_\mu \di_\nu D(\phiH) \comma \label{S2E5}
    \ee
    where $\wave \equiv \eta^{\mu \nu}\,\di_\mu \di_\nu$.

    Equation (\ref{S2E5}) has the solution
    $D(\phiH) = A\,t + B\,x + C$, where $A$, $B$ and $C$ are
    integration constants. If we require $D(\phiH) = $ constant
    and set this constant to be $D(0)$, it implies $\phiH = 0$
    \footnote{This procedure only simplifies the forthcoming
    calculation but it is not necessary. Similar calculations without
    making use of this procedure will be done in the type-II case.}
    such that the field equations (\ref{S2E1}) and (\ref{S2E2})
    simply become
    \be
      {\cJ} \b = \b
      \exp\!\left(\,\int^{\vphiH}\,\frac{H(u)}{D'(u)}\,\de
u\,\right)\,\left[\,D'(\vphiH)\,\hat{R} - \frac{H(\vphiH)}{D'(\vphiH)}\,\DelH^2
D(\vphiH)\,\right] \comma \B \label{S2E6}
    \ee
    \be
      \const\,\Ta_{\mu \nu} \b = \b
      \DelH^2 D(\vphiH)\,\gH_{\mu \nu} - \DelH_\mu\!\DelH_\nu D(\vphiH) \period
\label{S2E7}
    \ee

    If equations (\ref{S2E6}) and (\ref{S2E7}) are expanded about
    $\vphiH = 0$ and $\gH_{\mu \nu} = \eta_{\mu \nu}$, the linear
    parts of the expansions are
    \begin{eqnarray}
      \cJ & = &
      \minus \half\,D'(0)\,\wave \hH - H(0)\,\wave \vphiH \comma \label{S2E8}
\\
      \const\,\Ta_{\mu \nu} & = &
      D'(0)\,\wave \vphiH\,\eta_{\mu \nu}
      - D'(0)\,\di_\mu \di_\nu \vphiH \comma \label{S2E9}
    \end{eqnarray}
    where the harmonic coordinate conditions
    \begin{eqnarray}
      g^{\mu \nu}\,\Gamma_{\mu \nu}{}^\lambda \b = \b
      \exp\,\left(\,\int^{\Phi} \frac{H(u)}{D'(u)}\,\de
u\,\right)\,\,\left[\,\eta^{\mu \nu}\,\eta^{\lambda \rho}\,\hat{\Gamma}_{\mu
\nu \rho} + O(\hH^2_{\mu \nu})\,\right] \b = \b 0 \B \label{S2E10}
    \end{eqnarray}
    are used when $\hat{R}$ is expanded. In equation (\ref{S2E8}), the
    term $\hH$ is defined to be $\hH = \eta^{\mu \nu}\,\hH_{\mu \nu}$.
    If equation (\ref{S2E9}) is contracted with the inverse flat
    metric $\eta^{\mu \nu}$, the resultant scalar equation will be
    \begin{eqnarray}
      \const\,\left(\,\Ta_{11} - \Ta_{00}\,\right) & = &
      D'(0)\,\wave \vphiH \label{S2E11}
    \end{eqnarray}
    which has a general solution
    \begin{eqnarray}
      \vphiH(t,x) & = &
      A_{\pm}(t \pm x)
      \mp \frac{4\,\pi\,G}{D'(0)}\,\int^{\infty}_{\minus \infty} \de u
\int^t\de v\,\Ta_{00}(v\pm|\,x-u\,|,u) \nonumber \\ &&
      \pm \s \frac{4\,\pi\,G}{D'(0)}\,\int^{\infty}_{\minus \infty} \de u
\int^t\de v\,\Ta_{11}(v\pm|\,x-u\,|,u) \comma \label{S2E12}
    \end{eqnarray}
    where $A_{\pm}$ are two arbitrary functions. On the other hand, by
    using (\ref{S2E11}), equation (\ref{S2E8}) implies that $\hH$ is
    given by
    \begin{eqnarray}
      \hH(t,x) & = &
      B_{\pm}(t \pm x)
      \pm \frac{\const\,H(0)}{{D'(0)}^2}\,\int^{\infty}_{\minus \infty} \de u
\int^t\de v\,\Ta_{00}(v\pm|\,x-u\,|,u) \nonumber \\ &&
      \mp \s \frac{\const\,H(0)}{{D'(0)}^2}\,\int^{\infty}_{\minus \infty} \de
u \int^t\de v\,\Ta_{11}(v\pm|\,x-u\,|,u) \nonumber \\ &&
      \mp \s \frac{1}{D'(0)}\,\int^{\infty}_{\minus \infty} \de
      u \int^t\de v\,{\cJ}(v\pm|\,x-u\,|,u) \comma \label{S2E13}
    \end{eqnarray}
    where $B_{\pm}$ are another two arbitrary functions. The
    components of $\hH_{\mu \nu}$ can be expressed in terms of $\hH$
    because the harmonic coordinate condition (\ref{S2E10}) is
    assumed. As a result, the original metric $g_{\mu \nu}$ can be
    computed.

    If equation (\ref{S2E9}) is expanded component-wise, we find that
    equation (\ref{S2E12}) is a general solution only when
    \begin{eqnarray}
      \dit \Ta_{00} \b = \b \dix \Ta_{10}, & \B {\rm and} \B &
      \dit \Ta_{10} \b = \b \dix \Ta_{11} \label{S2E14}
    \end{eqnarray}
    hold. However, we always have these conditions because one can
    show that (\ref{S2E14}) is just the first order expansion of
    the conservation laws (\ref{CL1}). Note that the solution
    (\ref{S2E13}) forces $c\epsilon = 0$ (as is easily seen by
    inserting the trace of (\ref{S2E5}) into (\ref{S2E4})) regardless
    of whether the stress-energy-momentum tensor obeys the local
    conservation laws or not.

  \subsection{Type II Equations} \label{WFA.2}
    We consider here the field equations (\ref{SFE2}) and (\ref{TFE2}).
    It is still convenient to perform the conformal transformation
    before carrying out the weak-field expansion, so we take
    \begin{eqnarray*}
      g_{\mu \nu} & = &
      \EXP^{\minus k\,\Phi}\,\gH_{\mu \nu} \comma
    \end{eqnarray*}
    and  the field equations (\ref{SFE2}) and (\ref{TFE2}) give
    \begin{eqnarray}
      2\,\Jb & = & D'(\Phi)\,\hat{R} - k\,\DelH^2 D(\Phi) \comma \label{S2E15}
\\
      \const\,\Tb_{\mu \nu} & = &
      \frac{1}{k}\,D'(\Phi)\,\hat{R}_{\mu \nu}
      - \DelH_\mu\!\DelH_\nu D(\Phi) + \half\,\DelH^2 D(\Phi)\,\gH_{\mu \nu}
\period \label{S2E16}
    \end{eqnarray}
    Furthermore, the dilaton source $\Jb$, conformal metric
    $\gH_{\mu \nu}$ and characteristic function $D(\Phi)$ are taken
    to have the form
    \begin{eqnarray*}
      \Jb \b = \b \epsilon\,c + {\cJ}  \comma \B
      \gH_{\mu \nu} \b = \b \eta_{\mu \nu} + \hH_{\mu \nu} & {\rm and} &
      D(\Phi) \b = \b \sigma_o + \sigma \comma
    \end{eqnarray*}
    where $\sigma_o$ and $\eta_{\mu \nu}$ are the vacuum solutions of
    (\ref{S2E15}) and (\ref{S2E16}). Therefore $\sigma_o$ has a
    general solution
    \begin{eqnarray}
      \sigma_o(t,x) & = &
      \frac{\epsilon\,c}{2\,k}\,\left(\,t^2 - x^2\,\right)
      + A\,t + B\,x + C \comma \label{S2E17}
    \end{eqnarray}
    where $A$, $B$ and $C$ are constants.

    Expanding equations (\ref{S2E15}) and (\ref{S2E16}) about
    $D(\Phi) = \sigma_o$ and $\gH_{\mu \nu} = \eta_{\mu \nu}$, and
    ignoring the non-linear and mixed terms of $\sigma$ and
    $\hH_{\mu \nu}$, the two equations become
    \begin{eqnarray}
      \minus 2\,\cJ & = & k\,\wave \sigma
      + k\,\hH^{\mu \nu}\,\di_\mu \di_\nu \sigma_o
      + \half\,D'(D\invs (\sigma_o))\,\wave \hH \comma \label{S2E18} \\
      \const\,\Tb_{\mu \nu} & = &
      \minus \frac{1}{4\,k}\,D'(D\invs (\sigma_o))\,\wave \hH\,\eta_{\mu \nu}
      - \di_\mu \di_\nu \sigma
      + \di_\lambda \sigma_o\,\hat{\Gamma}_{\mu \nu \rho}\,\eta^{\lambda \rho}
\nonumber \\ &&
      + \s \half\,\hH^{\lambda \rho}\,\di_\lambda\,\di_\rho \sigma_o\,\eta_{\mu
\nu}
      + \half\,\wave \sigma_o\,\hH_{\mu \nu}
      + \half\,\wave \sigma\,\eta_{\mu \nu} \comma \label{S2E19}
    \end{eqnarray}
    where the vacuum solutions are used and $\hH^{\lambda \rho}$ is
    the first order term in the expansion of $\gH^{\lambda \rho}$
    with respect to $\hH_{\mu \nu}$. As a useful tool for simplifying
    the calculation, harmonic coordinate conditions (\ref{S2E10}) have
    again been used. Since we can write
    \begin{eqnarray*}
      \di_\mu \di_\nu \sigma_o \b = \b
      \minus \frac{\epsilon\,c}{k}\,\eta_{\mu \nu} \qquad & {\rm and} & \qquad
      \hH^{\mu \nu} \b = \b
      \minus \eta^{\mu \alpha}\,\eta^{\nu \beta}\,\hH_{\alpha \beta} \comma
    \end{eqnarray*}
    equations (\ref{S2E18}) and (\ref{S2E19})  have the general solution
    \begin{eqnarray}
      \hH(t,x) & = &
      E_{\pm}(t \pm x)
      \pm \const\,k\,\int^{\infty}_{\minus \infty} \de u \int^t\de
v\,\frac{\Tb_{00}(v\pm|\,x-u\,|,u)}{D'(D\invs (\sigma_o(v\pm|\,x-u\,|,u)))} \B
\nonumber \\ &&
      \mp \s \const\,k\,\int^{\infty}_{\minus \infty} \de u \int^t\de
v\,\frac{\Tb_{11}(v\pm|\,x-u\,|,u)}{D'(D\invs (\sigma_o(v\pm|\,x-u\,|,u)))}
\comma \label{S2E20} \\
      \sigma(t,x) & = &
      F_{\pm}(t \pm x)
      \mp 4\,\pi\,G\,\int^{\infty}_{\minus \infty} \de u \int^t\de
v\,\Tb_{00}(v\pm|\,x-u\,|,u) \nonumber \\ &&
      \pm \s 4\,\pi\,G\,\int^{\infty}_{\minus \infty} \de u \int^t\de
v\,\Tb_{11}(v\pm|\,x-u\,|,u) \nonumber \\ &&
      \mp \s \frac{1}{k}\,\int^{\infty}_{\minus \infty} \de u \int^t\de
v\,{\cJ}(v\pm|\,x-u\,|,u) \nonumber \\ &&
      \mp \s \frac{\epsilon\,c}{2\,k}\,\int^{\infty}_{\minus \infty} \de u
\int^t\de v\,\hH(v\pm|\,x-u\,|,u) \comma \label{S2E21}
    \end{eqnarray}
    where $E_{\pm}$ and $F_{\pm}$ are four arbitrary functions. By
    using the harmonic coordinate conditions (\ref{S2E10}), equations
    (\ref{S2E17}), (\ref{S2E20}) and (\ref{S2E21}), the components of
    the original metric $g_{\mu \nu}$ can be found.

    Similar to the case we encountered in Section~\ref{WFA.1}, the
    solutions (\ref{S2E20}) and (\ref{S2E21}) only satisfy the
    tensorial field equation (\ref{S2E19}) when the (weak-field)
    covariant conservation law (\ref{CL2}) is satisfied.


\section{Post-Newtonian Approximation} \label{PNA}
  \bigskip

  Consider a system of particles that are bound together by their
  mutual gravitational attraction. Let $\bar{M}$, $\bar{r}$ and
  $\bar{v}$ be the typical values of the masses, separations and
  velocities of these particles. Since the typical kinetic energy
  $\bar{M}\,\bar{v}^2/2$ is roughly of the same order of magnitude
  as the typical gravitational potential energy
  $G\,\bar{M}^2\,\bar{r}$ in Newtonian mechanics, we have
  $\bar{v}^2 \sim G\,\bar{M}\,\bar{r}$ \cite{Mann1}.

  If a Schwarzschild-like coordinates are employed, $g_{\mu \nu}$
  can be expressed as \cite{Ross}
  \begin{eqnarray*}
    g_{00} \b = \b \minus 1 / g_{11} \b = \b
    \minus 1\,+ \gttB + \stackrel{\scriptscriptstyle 4}{g}_{00} + \cdots & \b
{\rm and} \b &
    g_{10} = \stackrel{\scriptscriptstyle 3}{g}_{10} +
\stackrel{\scriptscriptstyle 5}{g}_{10} + \cdots \comma
  \end{eqnarray*}
  where $\stackrel{\scriptscriptstyle N}{g}_{\mu \nu}$ is of order
  $\bar{v}^N$. Now if we write $\Phi(t,x)$ as
  $\Phi =\,\PhiO + \PhiA + \PhiB + \cdots$, where
  $\stackrel{\scriptscriptstyle N}{\Phi}$ is of order $\bar{v}^N$,
  the geometric parts of the field equations ${\cal J}$ and
  ${\cal T}_{\mu \nu}$ (i.e. the right sides of either the pair
  (\ref{SFE1}) and (\ref{TFE1}) of type-I equations or (\ref{SFE2})
  and (\ref{TFE2}) of the type-II equations) can be expanded
  \footnote{All the expansions in this section are done by using
  Maple V.}
  as
  \begin{eqnarray*}
    {\cal J} \b = \b \cJO + \cJA + \cJB \cdots & \B {\rm and} \B &
    {\cal T}_{\mu \nu} \b = \b
    \TO_{\mu \nu} + \TA_{\mu \nu} + \TB_{\mu \nu} + \cdots \comma
  \end{eqnarray*}
  where the terms $\stackrel{\scriptscriptstyle N}{\cal J}$ and
  $\stackrel{\scriptscriptstyle N}{\cal T}_{\mu \nu}$ are of order
  $\bar{v}^N/\bar{r}^2$.

  Simple arguments based on time-reversal invariance
  \cite{Mann1,Ross,Weinberg}, however,  imply that the
  stress-energy-momentum tensor $T_{\mu \nu}$ and the dilaton source
  ${J}$ have the expansions
  \begin{eqnarray*}
    J & = & \JO + \JB + \cdots \comma \\
    T_{00} \b = \b \TttO + \TttB + \cdots \comma \B
    T_{10} & = & \TtxA + \cdots \B {\rm and} \B
    T_{11} \b = \b \TxxB + \cdots \comma
  \end{eqnarray*}
  where $\stackrel{\scriptscriptstyle N}{J}$ has an order
  $\bar{v}^N/\bar{r}^2$ and
  $\stackrel{\scriptscriptstyle N}{T}_{\mu \nu}$ denotes the term of
  order $\bar{v}^{N+2}/\bar{r}^2$. Therefore we obtain the following
  relationships by matching the order of individual terms:
  \begin{eqnarray*}
    \cJO \b = \b \JO \comma \qquad
    \TO_{00} \b = \b \TO_{11} \b = \b 0 \comma \qquad
    \TB_{00} \b = \b \const\,\TttO \comma \quad \cdots
  \end{eqnarray*}
  and so on.

  \subsection{Type-I Equations} \label{PNA.1}
    Let us consider the zeroth order expansion of ${\cal J}$ and
    ${\cal T}_{\mu\nu}$ of the type-I field equations first. Expanding
    (\ref{SFE1}) and (\ref{TFE1}) we obtain
    \begin{eqnarray}
      \cJO & = &
      \minus 2\,H(\PhiO)\,\dixx\!\PhiO -
H'(\PhiO)\,\left(\,\dix\!\PhiO\,\right)^2 \comma \nonumber \\
      \TO_{00} & = &
      \half\,H(\PhiO)\,\left(\,\dix\!\PhiO\,\right)^2
      - D'(\PhiO)\,\dixx\!\PhiO
      - D''(\PhiO)\,\left(\,\dix\!\PhiO\,\right)^2 \comma \B \label{S3E1} \\
      \TO_{10} & = & 0 \comma \nonumber \\
      \TO_{11} & = &
      \half\,H(\PhiO)\,\left(\,\dix\!\PhiO\,\right)^2 \period \label{S3E2}
    \end{eqnarray}
    Since $\TO_{11}$ is zero, there are two possible cases:
    \begin{itemize}
      \item[I]\hspace{5mm}  $\dix\!\PhiO \b = \b 0 \comma$
      \item[II]\hspace{5mm} $\dix\!\PhiO \b \neq \b 0$ \B and \B
                            $H(\PhiO\!\!(t,x)) \b = \b 0 \period$
    \end{itemize}
    In both cases, $\cJO$ vanishes and so it is necessary for $\JO$ to
    be zero. In the next two parts, we shall consider these two cases
    separately.

    \subsubsection{Case I: $\dix\!\PhiO\,= 0$} \label{PNA.1.1}
      After we impose the condition $\dix\!\PhiO\,= 0$ on the first
      order expansions, the non-trivial expansions are
      \begin{eqnarray*}
	\cJA \b = \b \minus 2\,H(\PhiO)\,\dixx\!\PhiA & \B {\rm and} \B &
	\TA_{00} \b = \b \minus D'(\PhiO)\,\dixx\!\PhiA \period
      \end{eqnarray*}
      As $\cJA$ and $\TA_{00}$ must be zero, we conclude that
      $$
	\PhiA\!\!(t,x) \b = \b \PhiA_a\!\!(t)\,x \period
      $$

      If we now expand to second order with the aid of the restrictions
      $\dix\!\PhiO\,= 0$ and $\dixx\!\PhiA\,= 0$, we find
      \be
	\JB = 2 H(\PhiO) \left( \ditt\!\PhiO - \dixx\!\PhiB \right)
	+ H'(\PhiO) \left[ (\dit\!\PhiO)^2 - (\dix\!\PhiA)^2 \right]
	+ D'(\PhiO) \dixx\!\gttB \comma \label{S3E3}
      \ee
      \be
	\const\,\TttO =
	\half H(\PhiO) \left[ (\dit\!\PhiO)^2 + (\dix\!\PhiA)^2 \right]
	- D'(\PhiO) \dixx\!\PhiB
	- D''(\PhiO) (\dix\!\PhiA)^2 \comma \label{S3E4}
      \ee
      \be
	0 = H(\PhiO)\,\dit\!\PhiO\,\dix\!\PhiA
	- D'(\PhiO)\,\ditx\!\PhiA
	- D''(\PhiO)\,\dit\!\PhiO\,\dix\!\PhiA \comma \B \label{S3E5}
      \ee
      \be
	0 =
	\half\,H(\PhiO)\,\left[\,(\dit\!\PhiO)^2 + (\dix\!\PhiA)^2\,\right]
	- \ditt D(\PhiO) \period \label{S3E6}
      \ee
      Equations (\ref{S3E5}) and (\ref{S3E6}) can be integrated and give
      \begin{eqnarray}
	\PhiA_a\!\!(t) & = &
	A\,\exp\left(\,2\,\Lambda(\PhiO\!\!(t))\,\right) / D'(\PhiO\!\!(t)) \comma
\label{S3E7} \\
	\Dt D(\PhiO\!\!(t)) & = &
	\pm \sqrt{A^2\,\exp\left(\,4\,\Lambda(\PhiO\!\!(t))\,\right) +
B\,\exp\left(\,2\,\Lambda(\PhiO\!\!(t))\,\right)} \label{S3E8}
      \end{eqnarray}
      if $\dix\!\PhiA\,\neq 0$, where $A$ and $B$ are integration
      constants and $\Lambda$ is defined as
      \begin{eqnarray}
	\Lambda(\PhiO) & := &
	\half\,\int^{\PhiO} \frac{H(u)}{D'(u)}\,\de u \period \label{S3E9}
      \end{eqnarray}
      Here the term $D'(\PhiO\!\!(t))$ denotes the derivative of the
      function $D(\zeta)$ with respect to its argument $\zeta$ and is
      evaluated at $\zeta =\,\PhiO\!\!(t)$. Thus a term like
      $(1/D')'(\PhiO)$, which we shall encounter later, denotes
      $\minus D''(\PhiO) / [D'(\PhiO)]^2$. Therefore if $D$ and $H$ and
      the initial conditions are given, $\PhiO$ can be solved by using
      the differential equation (\ref{S3E8}). Once $\PhiO$ is known,
      $\PhiA_a$ can be found explicitly by using equation (\ref{S3E7}).
      However, it is not reasonable for $\PhiO$ and $\PhiA_a$ depend
      solely on $D$ and $H$ rather than the source. That is to say even
      if the spacetime is sourceless, we shall obtain non-static $\PhiO$
      and $\PhiA_a$. Thus we restrict ourselves to the case
      \begin{eqnarray*}
	\PhiO \b = \b \PhiO_o \b = \b {\rm constant} & \B {\rm and} \B &
	\PhiA_a\!\!(t) \b = \b 0 \period
      \end{eqnarray*}
      After some manipulation of equations (\ref{S3E3}) and
      (\ref{S3E4}), $\dixx\!\PhiB$ and $\dixx\!\gttB$ can be written
      down explicitly and they yield
      \begin{eqnarray*}
	\PhiB\!\!(t,x) & = &
	\minus \frac{4\,\pi\,G}{D'(\PhiO_o)}\,\int^{\infty}_{\minus \infty} |\,x -
u\,|\,\TttO\!\!(t,u)\,\de u \comma \\
	\nonumber \\
	\gttB\!\!(t,x) & = &
	\frac{1}{2\,D'(\PhiO_o)}\,\int^{\infty}_{\minus \infty} |\,x - u
\,|\,\JB\!\!(t,u)\,\de u \nonumber \\ &&
	- \s
\const\,\frac{H(\PhiO_o)}{\left[\,D'(\PhiO_o)\,\right]^2}\,\int^{\infty}_{\minus \infty} |\,x - u \,|\,\TttO\!\!(t,u)\,\de u \period
      \end{eqnarray*}
      Straightly speaking, there are arbitrary integration functions of
      $t$ in the solutions above ; adopting static boundary conditions at large
$x$
     renders them to be constants. Moreover, those integration terms
      linear in $x$, which come from integration with respect to $x$
      twice, are ignored because when the spacetime is sourceless, such
      term withs constant coefficients must be zero in order to achieve
      vacuum state.

      The non-trivial third order field equations give
      \begin{eqnarray*}
        \minus D'(\PhiO_o)\,\ditx \PhiB\!\!(t,x) & = &
        \const\,\TtxA\!\!(t,x) \comma \\
	\dixx \PhiC\!\!(t,x) & = & 0 \period
      \end{eqnarray*}
      It is obvious that the last equation implies
      $\PhiC\!\!(t,x)\,= 0$  by the aforementioned boundary conditions and the
first
      one is compatible with the $t$-component of the third order
      expansion of the conservation law
      \begin{eqnarray*}
        \dix \TtxA\!\!(t,x) & = & \dit \TttO\!\!(t,x) \period
      \end{eqnarray*}

      In the ${\rm n}^{\rm th}$ order expansions, the highest order
      second derivative terms will be
      $\dixx\!\stackrel{\scriptscriptstyle n}{\Phi}$,
      $\dixx\!\stackrel{\scriptscriptstyle n}{g}_{00}$,
      $\ditx\!\stackrel{\scriptscriptstyle n-1}{\Phi}$,
      $\ditx\!\stackrel{\scriptscriptstyle n-1}{g}_{10}$,
      $\ditt\!\stackrel{\scriptscriptstyle n-2}{\Phi}$ and
      $\ditt\!\stackrel{\scriptscriptstyle n-2}{g}_{00}$ if $n$ is even.
      When $n$ is odd, the derivatives are
      $\dixx\!\stackrel{\scriptscriptstyle n}{\Phi}$,
      $\dixx\!\stackrel{\scriptscriptstyle n}{g}_{10}$,
      $\ditx\!\stackrel{\scriptscriptstyle n-1}{\Phi}$,
      $\ditx\!\stackrel{\scriptscriptstyle n-1}{g}_{00}$,
      $\ditt\!\stackrel{\scriptscriptstyle n-2}{\Phi}$ and
      $\ditt\!\stackrel{\scriptscriptstyle n-2}{g}_{10}$.
      In either case, there are, at most, six unknown derivatives but
      only four equations from the field equations are available. As a
      result, we have to fix the gauge freedom in order to obtain a
      closed system. In our case, the metric $g_{\mu \nu}$ depends on
      two functions: $g_{00}$ and $g_{10}$. As a result, there is one
      more degree of freedom in the coordinate choice. If the
      $t$-component of the harmonic coordinate conditions is used here,
      that is to say
      \begin{eqnarray*}
        \eta^{\mu \nu}\,\Gamma_{\mu \nu}{}^0 =
        \left[\,\dit \gttB - \dix \gtxC\,\right] + O(\bar{v}^5) = 0 \comma
      \end{eqnarray*}
      there will be two more equations to describe the second
      derivatives of the metric because the harmonic coordinate
      conditions involve only first derivatives of the metric. Therefore
      after the use of the gauge above, one can find that the fourth
      order expansions of the $t$-$x$-component of the metric field
      equation is trivial but the dilaton and $t$-$t$-component of
      the metric equation give
      \begin{eqnarray}
	\dixx \gttD\!\!(t,x) & = &
	\ditt \gttB\!\!(t,x)
	+ \left( \frac{1}{D'} \right)'(\PhiO_o)\,\PhiB\!\!(t,x)\,\JB\!\!(t,x)
	+ \frac{\JD\!\!(t,x)}{D'(\PhiO_o)} \nonumber \\ &&
	+ \s
\frac{16\,\pi\,G\,H(\PhiO_o)}{\left[\,D'(\PhiO_o)\,\right]^2}\,\left[\,\TxxB\!\!(t,x) - \TttB\!\!(t,x) - \gttB\!\!(t,x)\,\TttO\!\!(t,x)\,\right] \nonumber \\ &&
	- \s 16\,\pi\,G\,\left( \frac{H}{\left[\,D'\,\right]^2}
\right)'(\PhiO_o)\,\PhiB\!\!(t,x)\,\TttO\!\!(t,x) \nonumber \\ &&
        + \s
\frac{H'(\PhiO_o)\,D'(\PhiO_o)-2\,H(\PhiO_o)\,D''(\PhiO_o)}{\left[\,D'(\PhiO_o)\,\right]^2}\,\left[\,\dix \PhiB\!\!(t,x)\,\right]^2 \comma \label{S3E10} \\
	\dixx \PhiD\!\!(t,x) & = &
	\minus \frac{\const}{D'(\PhiO_o)}\,\TttB\!\!(t,x)
	- \frac{16\,\pi\,G}{D'(\PhiO_o)}\,\gttB\!\!(t,x)\,\TttO\!\!(t,x) \nonumber \\
&&
	- \s \const\,\left( \frac{1}{D'}
\right)'(\PhiO_o)\,\PhiB\!\!(t,x)\,\TttO\!\!(t,x) \nonumber \\ &&
	+ \frac{H(\PhiO_o)-2\,D''(\PhiO_o)}{2\,D'(\PhiO_o)}\,\left[\,\dix
\PhiB\!\!(t,x)\,\right]^2 \nonumber \\ &&
	+ \half\,\dix \gttB\!\!(t,x)\,\dix \PhiB\!\!(t,x) \period \label{S3E11}
      \end{eqnarray}
      Equations (\ref{S3E10}) and (\ref{S3E11}) can be used to find
      $\gttD\!\!(t,x)$ and $\PhiD\!\!(t,x)$. There is one more equation
      from the expansions but one can show that it is equivalent to the
      $x$-component of the conservation laws.

    \subsubsection{Case II: $\dix\!\PhiO\,\neq 0$} \label{PNA.1.2}
      When this is the case, $H(\PhiO)$ and $\dixx D(\PhiO)$ must be
      zero according to (\ref{S3E1}) and (\ref{S3E2}). Since
      $H(\PhiO) = 0$, all the spatial derivatives of $H(\PhiO)$ must
      be zero and so $H^{(n)}(\PhiO) = 0$ for all $n$ because
      $\dix\!\PhiO$ is non-zero. Thus, the $H$-dependence of the
      expansions is gone because $H(\Phi)$ is Taylor expanded about
      $\Phi =\,\PhiO$.

      The nontrivial first order expansions will be as follows if the
      restrictions above are imposed:
      \begin{eqnarray*}
        \TA_{00} \b = \b
        \minus \dixx \left[\,\PhiA\,D'(\PhiO)\,\right] & \B {\rm and} \B &
        \TA_{10} \b = \b \minus \ditx D(\PhiO) \period
      \end{eqnarray*}
      Since $\TA_{10}\,= 0$, the term $\dix D(\PhiO)$ has a constant
      value $A$ because $\dixx D(\PhiO) = 0$ from the zeroth order
      expansion. Therefore we have
      \begin{eqnarray}
        D(\PhiO\!\!(t,x)) & = & A\,x + A_a(t) \comma \label{S3E12}
      \end{eqnarray}
      where $A_a(t)$ is an arbitrary function of $t$.

      If the previous restrictions are imposed on the second order
      expansions, the expansions will give
      \begin{eqnarray}
        \cJB & = & \dixx\!\gttB\,D'(\PhiO) \comma \nonumber \\
        \TB_{00} & = & \half\,A\,\dix\!\gttB
        - \dixx \left[\,D'(\PhiO)\,\PhiB\,\right]
        - \half\,\dixx \left[\,D''(\PhiO)\,\PhiA^2\,\right] \comma \nonumber \\
        \TB_{10} & = & \minus \ditx \left[\,\PhiA\,D'(\PhiO)\,\right] \comma
\nonumber \\
        \TB_{11} & = & \minus \frac{A}{2}\,\dix\!\gttB - A_a''(t) \period
\label{S3E13}
      \end{eqnarray}
      Because $\TA_{00}\,=\,\TB_{10}\,= 0$, $\dix [ \PhiA\,D'(\PhiO) ]$
      is constant and we may write
      \begin{eqnarray*}
        \PhiA\!\!(t,x) & = &
        \left[\,B_1\,x + B_2\,\right] / D'(\PhiO\!\!(t,x)) \comma
      \end{eqnarray*}
      where $B_1$ and $B_2$ are constants of integration.

      As $\dix\!\PhiO\,\neq 0$, the constant $A$ in (\ref{S3E12}) cannot
      be zero. Thus (\ref{S3E13}) implies
      \begin{eqnarray*}
        \gttB\!\!(t,x) & = &
        \minus \frac{2}{A}\,A_a''(t)\,x + {\gttB}_b
      \end{eqnarray*}
      because $\TB_{11} = 0$. As a result, we are able to conclude that
      $\cJB\,= 0$. In other words, when $\JB\,\neq 0$, one is limited
      to the case $\dix \PhiO\!\!(t,x) = 0$ as the only possible choice.

      Now if we use the $x$-component of the harmonic gauge
      $\dix\!\gttB = 0$, the second derivative of $A_a(t)$ can be
      removed. In other words, we have $A_a(t) = A_1\,t + A_2$ where
      $A_1$ and $A_2$ are some constants and we obtain
      \begin{eqnarray*}
        \PhiO\!\!(t,x) & = & D^{-1}(A\,x + A_1\,t + A_2) \comma \\
        \PhiB\!\!(t,x) & = &
        \minus
\frac{D''(\PhiO\!\!(t,x))}{2\,D'(\PhiO\!\!(t,x))}\,\PhiA^2\!\!(t,x)
        \,+ \PhiB_a\!x\,+ \PhiB_b \nonumber \\ &&
        - \s \frac{4\,\pi\,G}{D'(\PhiO\!\!(t,x))}\,\int^{\infty}_{\minus
\infty} |\,x - u\,|\,\TttO\!\!(t,u)\,\de u \comma \\
        \gttB\!\!(t,x) & = & {\gttB}_b \period
      \end{eqnarray*}
      The aformentioned choice of static boundary conditions for large $x$ will
set $A_1=0$.

      Higher order expansions will not be done here because this case,
      which is equivalent to the requirement $H = 0$, is not really
      interesting at all. (For example, the forthcoming calculation
      shows that $H = 0$ is not a proper model to describe a ``star''
      in $(1+1)$ dimensions.

  \subsection{Type II Equations} \label{PNA.2}
    If the type-II field equations (\ref{SFE2}) and (\ref{TFE2}) are
    expanded as before, the zeroth order expansions of the right sides
    of them become
    \begin{eqnarray}
      \cJO & = &
      \minus \frac{k}{2}\,\exp\left(\,\minus k
\PhiO\,\right)\,\left[\,2\,D'(\PhiO)\,\dixx\!\PhiO +
D''(\PhiO)\,\left(\,\dix\!\PhiO\,\right)^2\,\right] \comma \B \label{S3E14} \\
      \TO_{00} & = &
      \half\,\left[\,k\,D'(\PhiO) -
D''(\PhiO)\,\right]\,\left(\,\dix\!\PhiO\,\right)^2 \comma \nonumber \\
      \TO_{10} & = & 0 \comma \nonumber \\
      \TO_{11} & = &
      \minus D'(\PhiO)\,\dixx\!\PhiO
      + \half\,\left[\,k\,D'(\PhiO) -
D''(\PhiO)\,\right]\,\left(\,\dix\!\PhiO\,\right)^2 \period \label{S3E15}
    \end{eqnarray}
    As we require $\TO_{00}$ and $\TO_{11}$ be zero, this devolves
    into two cases:
    \begin{itemize}
      \item[I:]  \hspace{7mm} $\PhiO\!\!(t,x) \b = \b \PhiO\!\!(t)$
      \item[II:] \hspace{7mm} $\dix\!\PhiO\!\!(t,x) \b \neq \b 0$ \comma
			      $k\,D'(\PhiO) \b = \b D''(\PhiO)$  \b and
			      \b $\dixx\!\PhiO\!\!(t,x) \b = \b 0$.
    \end{itemize}
    Let us consider these two possibilities separately.

    \subsubsection{Case I: $\dix \PhiO\!\!(t,x) = 0$} \label{PNA.2.1}
      Because $\PhiO\!\!(t,x) =\b\PhiO\!\!(t)$, the right sides of
      equations (\ref{S3E14}) to (\ref{S3E15}) will be zero and the
      zeroth order of the dilaton source $\JO$ must be zero again in
      this case.  Choosing boundary conditons so that $\PhiO$ is static at
large $x$ gives
   \begin{eqnarray*}
        \PhiO\!\!(t,x) & = & \PhiO_o \period
      \end{eqnarray*}

      If we consider the first order expansions, the non-trivial
      expansions will be
      \begin{eqnarray*}
	0 = \minus k\,\exp\left( \minus k \PhiO_o \right)\,D'(\PhiO_o)\,\dixx\!\PhiA &
\b {\rm and} \b &
	0 = \minus D'(\PhiO_o)\,\dixx\!\PhiA
      \end{eqnarray*}
      because $\cJA$ and $\TA_{11}$ must vanish. Since $D'$ is assumed
      to be non-zero, we obtain
      \begin{eqnarray*}
	\PhiA\!\!(t,x) & = & \PhiA_a\!x\ \period
      \end{eqnarray*}
      As a result, the non-trivial second order field equations imply
      \begin{eqnarray*}
        \JB & = &
        \minus \frac{k^2}{2}\,D'(\PhiO_o)\,\PhiA_a^2\,\exp\left(\,\minus k
\PhiO_o\,\right) \comma \\
	\dixx \gttB & = &
        k\,\left[\,k - \frac{D''(\PhiO_o)}{D'(\PhiO_o)}\,\right]\,\PhiA_a^2
	-\,\frac{16\,\pi\,G\,k}{D'(\PhiO_o)}\,\TttO(t,x) \comma \\
	\dixx \PhiB & = &
	\minus \frac{\JB\,\exp\left(\,k \PhiO_o\,\right)}{k\,D'(\PhiO_o)}
        + \left[\,\frac{k}{2} -
\frac{D''(\PhiO_o)}{D'(\PhiO_o)}\,\right]\,\PhiA_a^2
	-\,\frac{\const}{D'(\PhiO_o)}\,\TttO(t,x)
      \end{eqnarray*}
      because $\TB_{11}\,= 0$, $\cJB\,=\,\JB$ and
      $\TB_{00}\,= \const\!\TttO$ . Notice that in this case, the source
      $\JB$ must be a constant. By using these results, we found that
      the third order expansions simply imply
      \begin{eqnarray*}
	\PhiA_a & = & 0 \comma \\
	\PhiC\!\!(t,x) & = & \PhiC_o\!x \comma \\
	8\,\pi\,G\,\TtxA\!\!(t,x) & = &
	\minus D'(\PhiO_o)\,\ditx \PhiB\!\!(t,x) \period
      \end{eqnarray*}
      The last equation yields nothing new because it is compatible with
      the $t$-component of the conservation laws (\ref{CL2}). As
      $\PhiA_a\,= 0$, we have
      \begin{eqnarray*}
        \JB & = & 0 \comma \\
        \gttB\!\!(t,x) & = &
        \minus \frac{\const\,k}{D'(\PhiO_o)}\,\int^{\infty}_{\minus \infty}
|\,x - u\,|\,\TttO\!\!(t,u)\,\de u \comma \\
        \PhiB\!\!(t,x) & = &
        \minus \frac{4\,\pi\,G}{D'(\PhiO_o)}\,\int^{\infty}_{\minus \infty}
|\,x - u\,|\,\TttO\!\!(t,u)\,\de u \period
      \end{eqnarray*}

      Finally, by using the harmonic gauge $\dit\!\gttB\,= \dix\!\gtxC$,
      one can show that in the fourth order expansions, the
      $t$-$x$-component of metric equation is trivial, the dilaton
      equation is compatible with the $x$-component of the conservation
      laws and the other two expansions give
      \begin{eqnarray*}
        \dixx \gttD\!\!(t,x) & = &
        \ditt \gttB\!\!(t,x)
	- 2\,\frac{\exp\left(\,k\,\PhiO_o\,\right)}{D'(\PhiO_o)}\,\JD\!\!(t,x)
\nonumber \\ &&
        + \s \frac{16\,\pi\,G\,k}{D'(\PhiO_o)}\,\left[\,\TxxB\!\!(t,x) -
\TttB\!\!(t,x) - \gttB\!\!(t,x)\,\TttO\!\!(t,x)\,\right] \nonumber \\ &&
        + \s
16\,\pi\,G\,k\,\frac{D''(\PhiO_o)}{\left[\,D'(\PhiO_o)\,\right]^2}\,\PhiB\!\!(t,x)\,\TttO\!\!(t,x) \nonumber \\ &&
        - \s \frac{k\,D''(\PhiO_o)}{D'(\PhiO_o)}\,\left[\,\dix
\PhiB\!\!(t,x)\,\right]^2 \comma \\
        \dixx \PhiD\!\!(t,x) & = &
        \minus \frac{\const}{D'(\PhiO_o)}\,\TttB\!\!(t,x)
        - \frac{16\,\pi\,G}{D'(\PhiO_o)}\,\gttB\!\!(t,x)\,\TttO\!\!(t,x)
\nonumber \\ &&
        + \s
\const\,\frac{D''(\PhiO_o)}{\left[\,D'(\PhiO_o)\,\right]^2}\,\PhiB\!\!(t,x)\,\TttO\!\!(t,x) \nonumber \\ &&
	- \s \frac{\exp\left(\,k\,\PhiO_o\,\right)}{k\,D'(\PhiO_o)}\,\JD\!\!(t,x)
        + \left[\,\frac{k}{2} -
\frac{D''(\PhiO_o)}{D'(\PhiO_o)}\,\right]\,\left[\,\dix
\PhiB\!\!(t,x)\,\right]^2 \nonumber \\ &&
        + \s \half\,\dix \gttB\!\!(t,x)\,\dix \PhiB\!\!(t,x) \period
      \end{eqnarray*}
      Therefore $\gttD\!\!(t,x)$ and $\PhiD\!\!(t,x)$ are calculable in
      principle.

    \subsubsection{Case II: $\dix\!\PhiO\!\!(t,x) \neq 0$} \label{PNA.2.2}
      When this is the case, we have
      \begin{eqnarray}
	D(\PhiO) & = &
	A\,\exp\left(\,k \PhiO\,\right) + B \comma \label{S3E16} \\
	\PhiO & = &
	\PhiO_a\!x\,+ \PhiO_b \comma \label{S3E17}
      \end{eqnarray}
      where $A \neq 0$ and $B$ are constants and $\PhiO_a\,\neq 0$ and
      $\PhiO_b$ are arbitrary integration constants. As a result, if
      equations (\ref{S3E16}) and (\ref{S3E17}) hold, the only
      non-trivial zeroth order expansion gives
      $\JO\,= \minus A\,k^3 \PhiO_a^2 / 2$. In contrast to the other
      cases considered thus far, $\JO$ is here non-zero. If we
      substitute equations (\ref{S3E16}) and (\ref{S3E17}) into the
      first order expansions, one will easily conclude that
      \begin{eqnarray}
	\PhiA\!\!(t,x) & = & \PhiA_a \period \label{S3E18}
      \end{eqnarray}

      After we have imposed equations (\ref{S3E16}), (\ref{S3E17}) and
      (\ref{S3E18}) on the second order equation $\TB_{11}\,= 0$, it
      yields
      \begin{eqnarray}
	\dix\!\PhiB & = &
	\frac{1}{2\,k}\,\dix\!\gttB + \frac{\PhiO_a}{2}\,\gttB
        + \PhiB_a\!\!(t) \comma \label{S3E19}
      \end{eqnarray}
      where $\PhiB_a$ is an arbitrary function of $t$ which can be
      determined by the other two equations $\cJB\,=\,\JB$ and
      $\TB_{00}\b= \const\,\TttO$ as
      \begin{eqnarray*}
	\JB\!\!(t) & = & \minus k^3\,A\,\PhiO_a\,\PhiB_a\!\!(t) \comma \\
	\dixx\!\gttB\!\!(t,x) & = &
	\minus k \PhiO_a \dix\!\gttB
	- \frac{16 \pi G}{A} \exp\left( \minus k \PhiO \right) \TttO\!\!(t,x) \period
      \end{eqnarray*}
      Finally, the remaining equation $\TB_{10}\,= 0$ is just trivially
      satisfied. As a result, we can find the term $\gttB\!\!(t,x)$ if
      $\TttO\!\!(t,x)$ is known. Once we know $\gttB\!\!(t,x)$,
      $\PhiB\!\!(t,x)$ can be obtained by using equation (\ref{S3E19}).
      Since the restriction (\ref{S3E16}) is similar to the
      string-inspired field theory, reader may consult \cite{Ross} for
      similar third and fourth order expansions.



\section{Stellar Structure} \label{Stellar}
  \bigskip

  The existence of ``stars'' in two-dimensional spacetime is
  governed by the equation of hydrostatic equilibrium \cite{Weinberg}
  \begin{eqnarray}
    \minus \Dx p & = &
    (p + \rho)\,\Dx \ln(\sqrt{\minus g_{00}}) \comma \label{S4E1}
  \end{eqnarray}
  where $p$ is the pressure and $\rho$ is the density of the star.
  As we have two different sets of equations, we shall consider them
  separately. In this section, a static metric of the form
  \begin{eqnarray}
    \de s^2 & = & \minus B^2(x)\,\de t^2 + \de x^2 \label{S4E2}
  \end{eqnarray}
  with a perfect fluid stress tensor will be used to model the
  interior of a star in equilibrium state.

  \subsection{Type I Equations} \label{Stellar.1}
    If a perfect fluid is used, equation (\ref{S4E1}) implies that the
    dilaton source $\Ja$ is zero. Therefore the type-I field equations
    (\ref{SFE1}) and (\ref{TFE1}) can be written as
    \be
      \minus D'(\Phi)\,B''(x) =
      H(\Phi)\,\Dx \left[\,B(x)\,\Phi'(x)\,\right]
      + \frac{H'(\Phi)}{2}\,B(x)\,\left[\,\Phi'(x)\,\right]^2
      \comma \B \label{S4E3}
    \ee
    \be
      \const\,p(x) =
      \half\,H(\Phi)\,\left[\,\Phi'(x)\,\right]^2
      + \Dx \ln(B)\,\Dx D(\Phi) \comma \label{S4E4}
    \ee
    \be
      \const\,B(x)\,\left[\,p(x) - \rho(x)\,\right] =
      \Dx \left[\,B(x)\,\Dx D(\Phi)\,\right] \label{S4E5}
    \ee
    because the static condition is assumed.

    We first consider the sub-case when $H(\Phi) \neq 0$. Then the
    last two equations above can be rearranged algebraically such that
    $$
      \Phi'(x) =
      \minus \frac{D'(\Phi)}{H(\Phi)}\,\Dx \ln(B)
      \pm \sqrt{\left[\,\frac{D'(\Phi)}{H(\Phi)}\,\Dx \ln(B)\,\right]^2
      + \frac{16\,\pi\,G\,p}{H(\Phi)}} \comma
    $$
    $$
      \Phi''(x) =
      \const\,\frac{p - \rho}{D'(\Phi)}
      - \frac{16\,\pi\,G\,D''(\Phi)\,p}{D'(\Phi)\,H(\Phi)}
      + \frac{2\,D''(\Phi) - H(\Phi)}{H(\Phi)}\,\Dx \ln(B)\,\Phi'(x) \period
    $$
    Given the metric (\ref{S4E2}), equation (\ref{S4E1}) implies that
    \begin{eqnarray}
      \frac{B'(x)}{B(x)} = \minus \frac{p'(x)}{p + \rho} & \B {\rm and} \B &
      \frac{B''(x)}{B(x)} =
      \frac{2\,p'(x) + \rho'(x)}{(p + \rho)^2}\,p'(x)
      - \frac{p''(x)}{p + \rho} \period \label{S4E6}
    \end{eqnarray}
    Consequently, we obtain
    \be
      \Phi'(x) \b = \b
      \frac{D'(\Phi)}{H(\Phi)}\,\frac{p'(x)}{p + \rho}
      \pm \sqrt{\left[\,\frac{D'(\Phi)}{H(\Phi)}\,\frac{p'(x)}{p +
\rho}\,\right]^2 + \frac{16\,\pi\,G\,p}{H(\Phi)}} \comma \label{S4E7}
    \ee
    \be
      \Phi''(x) \b = \b
      \const\,\frac{p - \rho}{D'(\Phi)}
      - \frac{16\,\pi\,G\,D''(\Phi)\,p}{D'(\Phi)\,H(\Phi)}
      - \s \left[\,\frac{2\,D''(\Phi)}{H(\Phi)} - 1\,\right]\,\frac{p'(x)}{p +
\rho}\,\Phi'(x) \comma
    \ee
    \begin{eqnarray}
      \frac{B''(x)}{B(x)} & = &
      \const\,\left[\,2\,\frac{D''(\Phi)}{[D'(\Phi)]^2} -
\frac{H'(\Phi)}{H(\Phi)\,D'(\Phi)}\,\right]\,p
      - \const\,\frac{H(\Phi)}{[D'(\Phi)]^2}\,(p - \rho) \nonumber \\ &&
      - \s \frac{p'(x)}{p + \rho}\,\Dx
\ln\left(\,{H(\Phi)}{\left[\,D'(\Phi)\,\right]^2}\,\right) \period \label{S4E8}
    \end{eqnarray}
    If the equation of state $p = p(\rho)$ is given, (\ref{S4E7}) to
    (\ref{S4E8}) form a system of ordinary differential equations in
    $\Phi(x)$ and $\rho(x)$.

    In the Newtonian limit $p \approx 0$, $| p'(x) | \ll 1$ and
    $\Phi \approx 0$, we obtain
    \begin{eqnarray*}
      \frac{B''(x)}{B(x)}\,\rho(x) \approx p'(x)\,\Dx \ln(\rho) - p''(x)
      & \B {\rm and} \B &
      \frac{B''(x)}{B(x)} \approx
      \frac{\const\,H(0)}{[D'(0)]^2}\,\rho(x) \period
    \end{eqnarray*}
    As a result, if the fraction $[D'(0)]^2/H(0) = 2$, the results
    above will agree with Newton's equation of stellar equilibrium
    \begin{eqnarray}
      p'(x)\,\Dx \ln(\rho) - p''(x) & = & 4\,\pi\,G\,\rho^2(x) \label{NESE}
    \end{eqnarray}
    in two dimensions.

    When $H(\Phi) = 0$, equations (\ref{S4E3}) to (\ref{S4E5}) imply that
    \begin{eqnarray*}
      x\,p^\prime(x) + p(x) & = & \minus \rho(x) \period
    \end{eqnarray*}
    As $p(x)$ and $p^\prime(x)$ tend to zero in the Newtonian limit,
    the density $\rho(x)$ goes to zero in this limit. So $H = 0$ is
    not appropriate to model one-dimensional stellar structure.

  \subsection{Type II Equations} \label{Stellar.2}
    Consider the stellar structure with the type-II equations
    (\ref{SFE2}) and (\ref{TFE2}). Here the static metric and the
    stress tensor for a perfect fluid are used as before. The
    conservation of stress-energy implies via (\ref{CL2}) that $\Jb$
    is a constant. Therefore (\ref{S4E6}) still holds and the field
    equations can be expressed as
    \be
      \Phi'(x) =
      \frac{p'(x)}{k(p + \rho)}
      \pm \frac{1}{k}\,\sqrt{\left[\,\frac{p'(x)}{p + \rho}\,\right]^2 +
\frac{2}{D'(\Phi)}\,\left(\,\const\,k\,p - \Jb\,\EXP^{k\,\Phi}\,\right)}
      \comma \B \label{S4E9}
    \ee
    \be
      \Phi''(x) =
      \left[\,k -
\frac{D''(\Phi)}{D'(\Phi)}\,\right]\,\left[\,\Phi'(x)\,\right]^2
      + \frac{p'(x)}{p + \rho}\,\Phi'(x)
      - \frac{\const}{D'(\Phi)}\,(p + \rho) \comma
    \ee
    \be
      \frac{2\,p' + \rho'}{(p + \rho)^2}\,p'
      - \frac{p''}{p + \rho} =
      \frac{k}{2}\,\left[\,\frac{D''(\Phi)}{D'(\Phi)} - k\,\right]\,[\Phi']^2
      + \frac{k\,p'\,\Phi'}{p + \rho}
      + \frac{\const\,k}{D'(\Phi)}\,\rho \period \label{S4E10}
    \ee
    So that when the equation of state is given, equations
    (\ref{S4E9}) to (\ref{S4E10}) form a system of second order
    ordinary differential equations in $\Phi(x)$ and $\rho(x)$.

    If this solution is compared with Newton's equation of stellar
    equilibrium, in the Newtonian limit $\Jb \rightarrow 0$,
    $p \rightarrow 0$, $| p'(x) | \ll 1$ and $\Phi \rightarrow 0$,
    equation (\ref{S4E10}) will reduce to
    \begin{eqnarray*}
      p'(x)\,\Dx \ln(\rho) - p''(x) & \approx &
      \frac{\const\,k}{D'(0)}\,\rho^2(x) \period
    \end{eqnarray*}
    Thus this will agree with the Newtonian limit (\ref{NESE}) if
    $D'(0)$ equals $2\,k$.


\section{Cosmology} \label{Cosmo}
  \bigskip

  Let us consider the two-dimensional Robertson-Walker metric
  \begin{eqnarray*}
    \de s^2 & = &
    \minus \de t^2 + a^2(t)\,\frac{\de x^2}{1 - s\,x^2} \period
  \end{eqnarray*}
  Since the denominator of the $\de x^2$ term is only $x$-dependent,
  it can be removed by redefining the spatial coordinate. Although
  there are still three different cosmological models (closed, open
  and flat) the evolution of the scale factor $a(t)$ does not depend
  upon which model is under consideration, and we can take $s = 0$
  without loss of generality \cite{Mann1}. The metric then becomes
  \begin{eqnarray}
    \de s^2 & = & \minus \de t^2 + a^2(t)\,\de x^2 \period \label{S5E1}
  \end{eqnarray}
  Moreover, we assume a perfect fluid stress tensor with the
  equation of state
  \begin{eqnarray}
    p & = & ( \gamma - 1 )\,\rho \comma \label{EOS}
  \end{eqnarray}
  where $\gamma$ is a parameter.

  \subsection{Type-I Equations} \label{Cosmo.1}
    In this case, the field equations (\ref{SFE1}) and (\ref{TFE1}) read
    \begin{eqnarray}
      \rho(t) & = & \rho_o\,a^{\minus \gamma}(t) \comma \label{S5E2} \\
      a^{\gamma-1}(t)\,\Dt D(\Phi) & = &
      A\,\EXP^{(2-\gamma)\,\Lambda(\Phi)} \comma \label{S5E3} \\
      \const\,\rho_o\,\EXP^{(\gamma-1)\,\Lambda(\Phi)} & = &
      A\,\Dt \left[\,a(t)\,\EXP^{\Lambda(\Phi)}\,\right] \period \label{S5E4}
    \end{eqnarray}
    The functions $\Phi$ and $\rho$ are functions of time only,
    because the model is homogeneous. The constant $\rho_o$ comes
    from the conservation laws $\Del_\nu T^{\mu \nu} = 0$ and the
    equation of state (\ref{EOS}) \cite{Mann1}. The function $\Lambda$
    is defined in a manner identical to that of (\ref{S3E9}) in the
    study of the post-Newtonian approximation and $A$ is a non-zero
    integration constant. Although solving these equations for general
    $\gamma$ is difficult, solutions may easily be found for two
    special cases, $\gamma = 1$ and $\gamma = 2$, corresponding to
    dust- and radiation-filled universes, respectively.

    When $\gamma = 1$, equations (\ref{S5E3}) and (\ref{S5E4}) give
    \begin{eqnarray}
      a(t) & = &
      \left(\,\frac{\const\,\rho_o}{A}\,t + \alpha\,\right)\,\EXP^{\minus
\Lambda(\Phi)} \comma \label{S5E5} \\
      \Phi'(t) & = &
      A\,\EXP^{\Lambda(\Phi)} / D'(\Phi) \comma \label{S5E6}
    \end{eqnarray}
    where $\alpha$ is another integration constant. If the functions
    $D$ and $H$ are given, the dilaton field $\Phi(t)$ can be found
    from the first order (in general, nonlinear) ordinary differential
    equation (\ref{S5E6}). Once $\Phi$ is known, the density $\rho(t)$
    and the metric component $a(t)$ can be written down immediately as
    a consequence of equations (\ref{S5E2}) and (\ref{S5E5}).

    When $\gamma = 2$, the system (\ref{S5E3}) and (\ref{S5E4}) yields
    \begin{eqnarray}
      a(t) & = &
      \frac{A}{\beta}\,\exp\left(\,\frac{\const\,\rho_o}{A^2}\,D(\Phi) -
\Lambda(\Phi)\,\right) \comma \label{S5E7} \\
      \Phi'(t) & = &
      \frac{\beta}{D'(\Phi)}\,\exp\left(\,\Lambda(\Phi) -
\frac{\const\,\rho_o}{A^2}\,D(\Phi)\,\right) \comma \label{S5E8}
    \end{eqnarray}
    where $\beta$ is an integration constant. As a result, if the
    characteristic functions $D$ and $H$ are given, the dilaton field
    $\Phi(t)$ can be solved by using the first order differential
    equation (\ref{S5E8}). Thus the density and the metric component
    can be expressed as functions of $t$ by using equations
    (\ref{S5E2}) and (\ref{S5E7}). This procedure is similar to the
    one in the dust-filled universe case.

  \subsection{Type II Equations} \label{Cosmo.2}
    For the metric (\ref{S5E1}) and the perfect fluid stress tensor
    with equation of state (\ref{EOS}), the type-II equations
    (\ref{SFE2}) and (\ref{TFE2}) become
    \begin{eqnarray}
      \rho(t) & = & \rho_o\,a^{\minus \gamma}(t) \comma \label{S5E9} \\
      \const\,\rho(t) & = &
      \minus \frac{\Jb}{k}\,\EXP^{k\,\Phi} + D'(\Phi)\,\Dt \ln(a)\,\Phi'(t)
      + \frac{k}{2}\,D'(\Phi)\,\left[\,\Phi'(t)\,\right]^2
      \comma \B \B \label{S5E10} \\
      D'(\Phi)\,\Phi''(t) & = &
      \gamma\,\frac{\Jb}{k}\,\EXP^{k\,\Phi}
      - (\gamma-1)\,D'(\Phi)\,\Dt \ln(a)\,\Phi'(t) \nonumber \\ &&
      + \s \left[\,k\,(1-\frac{\gamma}{2})\,D'(\Phi) -
D''(\Phi)\,\right]\,\left[\,\Phi'(t)\,\right]^2 \period \label{S5E11}
    \end{eqnarray}
    According to equation (\ref{CL2}), the dilaton source $\Jb$ is a
    constant. As in the last subsection, solutions for general
    $\gamma$ are quite difficult to obtain, and we consider only
    $\gamma = 1$ and $\gamma = 2$.

    \subsubsection{Dust-filled universe} \label{Cosmo.2.1}
    When the parameter $\gamma = 1$, the system (\ref{S5E10}) and
    (\ref{S5E11}) has a solution
    $$
      a(t) \b = \b
      \const\,\rho_o\,\Sigma(\Phi(t))\,\EXP^{\minus \frac{k}{2}\,\Phi(t)}\,\int
\Sigma^{\minus 2}(\Phi(t))\,\de t \comma
    $$
    \be
      \Dt D(\Phi) \b = \b
      \Sigma(\Phi)\,\EXP^{\frac{k}{2}\,\Phi} \comma \label{S5E12}
    \ee
    where $\alpha$ is a constant and
    $\Sigma(\Phi) := \pm \sqrt{2\,\Jb\,D(\Phi) / k + \alpha}$.
    If $D(\Phi)$ is given, the ordinary differential equation
    (\ref{S5E12}) can be used to solve for $\Phi$. Thus we are able to
    compute the density $\rho(t)$ and the metric component $a(t)$.

    \subsubsection{Radiation-filled universe} \label{Cosmo.2.2}
    When $\gamma = 2$, the system (\ref{S5E10}) and (\ref{S5E11}) becomes
    \begin{eqnarray}
      \omega\,k\,\,a(t)\,\Dt D(\Phi) & = &
      \Jb\,a^2(t)\,\EXP^{k\,\Phi} + \zeta \comma \label{S5E13} \\
      \Dt \left[\,a(t)\,\EXP^{\frac{k}{2}\,\Phi}\,\right] & = &
      \omega\,\EXP^{\frac{k}{2}\,\Phi} \comma \label{S5E14}
    \end{eqnarray}
    where $\zeta$ and $\omega$ are integration constants. We shall
    divide the discussion up into two cases because equation
    (\ref{S5E13}) may either be linear or quadratic in $a(t)$,
    depending on whether or not $\Jb$ vanishes.

    When $\Jb = 0$, equations (\ref{S5E13}) and (\ref{S5E14}) give
    \begin{eqnarray}
      a(t) & = &
\frac{\zeta}{\omega\,\sigma\,k}\,\exp\left(\,\frac{\omega^2\,k}{\zeta}\,D(\Phi)
- \frac{k}{2}\,\Phi\,\right) \comma \nonumber \\
      \Dt D(\Phi) & = &
      \sigma\,\exp\left(\,\frac{k}{2}\,\Phi -
\frac{\omega^2\,k}{\zeta}\,D(\Phi)\,\right) \comma \label{S5E15}
    \end{eqnarray}
    where $\sigma$ is another integration constant.

    On the other hand, if $\Jb \neq 0$, the two equations
    (\ref{S5E13}) and (\ref{S5E14}) become
    \begin{eqnarray}
      a(t) & = &
      \frac{\omega\,k\,\Dt D(\Phi) \pm \sqrt{\left[\,\omega\,k\,\Dt
D(\Phi)\,\right]^2 - 4\,\zeta\,\Jb\,\EXP^{k\,\Phi}}}{2\,\Jb\,\EXP^{k\,\Phi}}
      \comma \nonumber \\
      2\,\omega\,\Jb\,\EXP^{\frac{k}{2}\,\Phi} & = &
      \Dt \left\{\,\omega\,k\,\Dt D(\Phi) \pm \sqrt{\left[\,\omega\,k\,\Dt
D(\Phi)\,\right]^2 - 4\,\zeta\,\Jb\,\EXP^{k\,\Phi}}\,\right\}
      \period \B \B \label{S5E16}
    \end{eqnarray}

    If the characteristic function $D(\Phi)$ is given, the
    differential equations (\ref{S5E15}) and (\ref{S5E16}) can be used
    to solve for the field $\Phi(t)$ in both cases. As a result, the
    density $\rho(t)$ and the metric component $a(t)$ can be found in
    principle.

  \subsection{Inflationary Universe} \label{Cosmo.3}
    The inflationary universe paradigm was developed to address the
    flatness, horizon and monopole problems \cite{Collins} that beset
    the standard $(3+1)$-dimen\-sional model of cosmology. In $(1+1)$
    dimensions the situation is somewhat different. Previous
    investigations of  $(1+1)$-dimensional cosmology within the
    context of the R=T theory \cite{Kevin} indicated that (a) the
    standard problems of $(3+1)$-dimensional cosmology  were absent
    due to the lack of structure in the lower-dimensional universe and
    (b) the mechanisms for realizing an exponentially inflating
    universe involved rather unconventional assumptions within the
    context of $(1+1)$-dimensional physics.

    In this section we consider the circumstances under which
    inflation could occur for an arbitrary dilaton theory of gravity.
    In the last two sections we showed how to solve for the metric
    unknown $a(t)$ given the functions $D(\Phi)$ and $H(\Phi)$. In
    this subsection we assume the function $a(t)$ varies exponentially
    with $t$, and we search for functions $H(\Phi)$ and $D(\Phi)$
    which can realize this.

    In order to allow the early universe to expand exponentially,
    there must be an outward pressure \cite{Kevin}
    \begin{eqnarray}
      p & = & \minus \rho \period \label{S5E17}
    \end{eqnarray}
    This will happen when the universe is trapped in a false vacuum
    phase.

    \subsubsection{Type I Equations} \label{Cosmo.3.1}
    When the equation of state becomes (\ref{S5E17}), the corresponding
    value of the parameter $\gamma$ is zero. Let us suppose that we
    know the characteristic function $H(\Phi)$ in advance; thus the
    other function $D(\Phi)$ is the object we are trying to find. If
    we suppose the metric component $a(t)$ has the form
    \begin{eqnarray*}
      a(t) & = &
      \frac{\const\,\rho_o}{\alpha\,\beta\,A}\,\EXP^{\beta\,t} \comma
    \end{eqnarray*}
    where $\beta$ is a positive constant and $\alpha$ is a non-zero
    constant such that $\alpha\,A$ is positive, equations (\ref{S5E2})
    to (\ref{S5E4}) will give
    \begin{eqnarray}
      \rho(t) & = & \rho_o \comma \nonumber \\
      \beta\,t & = & \Sigma(\Phi) \comma \label{S5E18}
    \end{eqnarray}
    \be
      D(\Phi) \b = \b
\frac{\const\,\rho_o}{\alpha\,\beta}\,\left[\,\frac{2\,\alpha}{\beta}\,\Sigma(\Phi) - \frac{B}{\beta}\,\EXP^{\minus \Sigma(\Phi)}\,\right]
      + C \comma \label{S5E19}
    \ee
    where $B$ and $C$ are integration constants and $\Sigma(\Phi)$ is
    defined as
    \begin{eqnarray}
      \Sigma(\Phi) & := &
      \ln\left(\,\alpha\,\EXP^{\minus 2\,\Lambda(\Phi)}
      \pm \sqrt{\left[\,\alpha\,\EXP^{\minus 2\,\Lambda(\Phi)}\,\right]^2 +
B\,\EXP^{\minus 2\,\Lambda(\Phi)}}\,\right) \period \label{S5E20}
    \end{eqnarray}

    Although the left side of equation (\ref{S5E19}) is the
    characteristic function $D(\Phi)$, the equation is actually an
    integral equation because the function $\Sigma(\Phi)$ involves
    the integration of $D'(\Phi)$. Differentiating equation
    (\ref{S5E20}) yields, after some manipulation,
    \begin{eqnarray}
      \frac{\alpha\,\beta^2}{4\,\pi\,G\,\rho_o} & = &
      \mp \frac{H(\Phi)}{\left[\,D'(\Phi)\,\right]^2}\,\frac{\left[\,2\,\alpha
+ B\,\EXP^{\minus \Sigma(\Phi)}\,\right]^2}{\sqrt{\alpha^2 +
B\,\EXP^{2\,\Lambda(\Phi)}}} \period \label{S5E21}
    \end{eqnarray}
    It is now clear that the sign of the square root in equation
    (\ref{S5E20}) is determined by the sign of $\alpha$ and the sign
    of the function $H(\Phi)$ which is then restricted to be either
    positive or negative definite. If we define
    \begin{eqnarray*}
      \omega(\Phi) & := &
      \left[\,\alpha^2 + B\,\EXP^{2\,\Lambda(\Phi)}\,\right]^{\frac{1}{4}}
      \comma
    \end{eqnarray*}
    the square root of equation (\ref{S5E21}) becomes
    \begin{eqnarray}
      \epsilon\,\sqrt{\mp \frac{\alpha\,\beta^2}{4\,\pi\,G\,\rho_o}\,H(\Phi)}
      & = &
      \left[\,\frac{\alpha}{\omega(\Phi)} \pm
\omega(\Phi)\,\right]\,\frac{H(\Phi)}{D'(\Phi)} \comma \label{S5E22}
    \end{eqnarray}
    where $\epsilon$ is either $1$ or $\minus 1$, depending on the
    choice of the second square root operation on equation (\ref{S5E21}).

    When $B = 0$, the function $\omega(\Phi)$ equals
    $\sqrt{\,|\,\alpha\,|\,}$. Since the choice of $\epsilon$ is
    arbitrary, we can conclude that
    \begin{eqnarray}
      D(\Phi) & = &
      \pm \sqrt{\frac{16\,\pi\,G\,\rho_o}{\beta^2}}\,\int^{\Phi} \sqrt{\minus
H(u)}\,\de u
      + C \label{S5E23}
    \end{eqnarray}
    when $\alpha$ is either positive or negative. Furthermore, we can
    also conclude that $H(\Phi)$ must be strictly negative for any
    non-zero arbitrary integration constant $A$, in order that the
    type-I equation possess an exponential solution. Finally, the
    dilaton field $\Phi$ can be solved as a function of $t$ by using
    equation (\ref{S5E23}) for $D(\Phi)$ together with equation
    (\ref{S5E18}).

    When $B \neq 0$, we obtain the equation
    ${\cal H}(\Phi) = {\cal G}(\omega)$, where
    \begin{eqnarray*}
      {\cal H}(\Phi) :=
      \epsilon\,\sqrt{\frac{\beta^2}{4 \pi G \rho_o}} \int \sqrt{\mp \alpha
H(\Phi)}\,\de \Phi & {\rm and} &
      {\cal G}(\omega) :=
      \pm 4 \omega + 4 \alpha \int \frac{\de \omega}{\omega^2 \mp \alpha}
      \period
    \end{eqnarray*}
    Therefore equation (\ref{S5E22}) implies
    \begin{eqnarray}
      D'(\Phi) & = &
      \pm \frac{H(\Phi)}{{\cal H}^\prime(\Phi)}\,\frac{\left[\,{\cal
G}\invs\!\left(\,{\cal H}(\Phi)\,\right)\,\right]^2 \pm \alpha}{{\cal
G}\invs\!\left(\,{\cal H}(\Phi)\,\right)} \period \label{S5E24}
    \end{eqnarray}

    If the function $H(\Phi)$ is strictly positive,
    \begin{eqnarray}
      {\cal G}(\omega) & = & \pm 4\,\omega
      +
\frac{4\,\alpha}{\sqrt{\,|\,\alpha\,|\,}}\,\arctan\left(\,\frac{\omega}{\sqrt{\,|\,\alpha\,|\,}}\,\right) \label{S5E25}
    \end{eqnarray}
    but when $H(\Phi)$ is strictly negative,
    \begin{eqnarray*}
      {\cal G}(\omega) & = &
      \pm 4\,\omega
      + \frac{2\,\alpha}{\sqrt{\,|\,\alpha\,|\,}}\,\ln\left(\,\frac{\omega -
\sqrt{\,|\,\alpha\,|\,}}{\omega + \sqrt{\,|\,\alpha\,|\,}}\,\right) \period
    \end{eqnarray*}
    In the case when $H(\Phi)$ is strictly negative, equation
    (\ref{S5E24}) will have no zero because
    $\pm \alpha = |\,\alpha\,|$ but when $H(\Phi)$ is strictly
    positive, (\ref{S5E24}) will vanish if
    \begin{eqnarray}
      {\cal H}(\Phi) & = &
      {\cal G}(\,\pm \sqrt{\,|\,\alpha\,|\,}\,) \period \label{S5E26}
    \end{eqnarray}
    As ${\cal G}(\omega)$ in (\ref{S5E25}) is well-defined at the
    points $\pm \sqrt{\,|\,\alpha\,|\,}$, there always exist a point
    $\Phi = \Phi_o$ such that condition (\ref{S5E26}) holds (because
    we have an arbitrary integration constant in ${\cal H}(\Phi)$).
    Thus, the assumption (\ref{AS1}) is violated in this case.
    Therefore $H(\Phi)$ can only be strictly negative and we have
    \begin{eqnarray}
      D(\Phi) & = &
      {\rm sgn}(\alpha)\,\int
\frac{H(\Phi)}{X^\prime(\Phi)}\,\frac{\left[\,Y\invs\!\left(\,X(\Phi)\,\right)\,\right]^2 + |\,\alpha\,|}{Y\invs\!\left(\,X(\Phi)\,\right)}\,\de \Phi \comma \label{S5E27}
    \end{eqnarray}
    where the functions $X$ and $Y$ are defined as
    \begin{eqnarray*}
      X(\Phi) & := &
      \pm \sqrt{\frac{\beta^2}{4\,\pi\,G\,\rho_o}}\,\int \sqrt{\minus
|\,\alpha\,|\,H(\Phi)}\,\de \Phi \comma \\
      Y(\omega) & := &
      {\rm sgn}(\alpha)\,4\,\omega +
\frac{2\,\alpha}{\sqrt{\,|\,\alpha\,|\,}}\,\ln\left(\,\frac{\omega -
\sqrt{\,|\,\alpha\,|\,}}{\omega + \sqrt{\,|\,\alpha\,|\,}}\,\right) \period
    \end{eqnarray*}
    Finally, we can find the dilaton field $\Phi$ as a function of $t$
    by using equation (\ref{S5E27}) for $D(\Phi)$ together with
    equation (\ref{S5E18}).

    \subsubsection{Using the type-II equations} \label{Cosmo.3.2}
    Suppose
    \begin{eqnarray*}
      a(t) & = & \alpha\,\EXP^{\beta\,t}
    \end{eqnarray*}
    when $\gamma = 0$, where $\alpha$ and $\beta$ are positive
    constants, the system (\ref{S5E9}) to (\ref{S5E11}) gives
    \begin{eqnarray}
      \rho(t) & = & \rho_o \comma \nonumber \\
      \Dt D(\Phi) & = &
      \frac{2}{k\,\beta\,A}\,\EXP^{\beta\,t + k\,\Phi} \comma \nonumber \\
      \const\,\rho_o\,k\,\beta\,A\,\EXP^{\beta\,t + \chi(t)} & = &
      \Dt \left[\,\EXP^{2\,\beta\,t + k\,\Phi}\,\EXP^{\chi(t)}\,\right]
      \comma \label{S5E28}
    \end{eqnarray}
    where $A$ is an integration constant and $\chi(t)$ is defined as
    \begin{eqnarray*}
      \chi(t) & := & \Jb\,A\,\EXP^{\minus \beta\,t} \period
    \end{eqnarray*}

    When the dilaton source $\Jb$ equals zero, the function $\chi(t)$
    vanishes. This simplification yields
    \begin{eqnarray}
      \Phi(t) & = &
      \frac{1}{k}\,\ln\left(\,\const\,\rho_o\,k\,A\,\EXP^{\minus \beta\,t} +
B_1\,\EXP^{\minus 2\,\beta\,t}\,\right) \comma \nonumber \\
      D(\Phi) & = &
      C_1 + \frac{16\,\pi\,G\,\rho_o}{\beta^2}\,\Sigma(\Phi)
      - \frac{2\,B_1}{k\,\beta^2\,A}\,\EXP^{\minus \Sigma(\Phi)}
      \comma \label{S5E29}
    \end{eqnarray}
    where $B_1$ and $C_1$ are integration constants and
    $\Sigma(\Phi)$ is defined as
    \begin{eqnarray*}
      \Sigma(\Phi) & := &
      \ln\left(\,4\,\pi\,G\,\rho_o\,k\,A\,\EXP^{\minus k\,\Phi} \pm
\sqrt{\left[\,4\,\pi\,G\,\rho_o\,k\,A\,\EXP^{\minus k\,\Phi}\,\right]^2 +
B_1\,\EXP^{\minus k\,\Phi}}\,\right) \period
    \end{eqnarray*}

    In the case when $\Jb$ is not zero, equation (\ref{S5E28}) becomes
    \begin{eqnarray*}
      \frac{\Jb}{\const\,\rho_o\,k}\,\EXP^{k\,\Phi} & = &
      \chi(t) - \chi^2(t)\,\Pi(\Phi)\,\EXP^{\minus \chi(t)}
      + B_2\,\chi^2(t)\,\EXP^{\minus \chi(t)} \comma
    \end{eqnarray*}
    where $B_2$ is another integration constant and the function
    $\Pi(\Phi)$ is defined as
    \begin{eqnarray*}
      \Pi(u) \b := \b \int^{u} \frac{1}{s}\,\EXP^{s}\,\de s \b = \b
      \ln(u) + \sum^{\infty}_{n=1} \frac{u^n}{n\,n!} \period
    \end{eqnarray*}
    Thus, we can write the function $D$ as
    \begin{eqnarray}
      D(\Phi) & = &
      \frac{16\,\pi\,G\,\rho_o}{\beta^2}\,\left[\,C_2 + B_2\,\EXP^{\minus
\theta(\Phi)} - \Pi(\theta(\Phi))\,\EXP^{\minus \theta(\Phi)}\,\right]
      \comma \label{S5E30}
    \end{eqnarray}
    where $C_2$ is a constant and $\theta(\Phi)$ is implicitly defined as
    \begin{eqnarray*}
      \frac{\Jb}{\const\,\rho_o\,k}\,\EXP^{k\,\Phi} & = &
      \theta + B_2\,\theta^2\,\EXP^{\minus \theta}
      - \Pi(\theta)\,\theta^2\,\EXP^{\minus \theta} \period
    \end{eqnarray*}

    The ordinary differential equations (\ref{S5E12}), (\ref{S5E15})
    or (\ref{S5E16}) can be solved completely in principle if the
    function $D$ is given. However, when the function $D$ in either
    equation (\ref{S5E29}) or (\ref{S5E30}) is used, all the
    differential equations mentioned above appear to be analytically
intractable, rendering
    numerical solutions the only practical option.


\section{Conclusions} \label{Conclus}
  \bigskip

   We have investigated many of the basic properties of a wide class of
$(1+1)$-dimensional dilaton theories
   of gravity coupled to matter.  In this section we recapitulate our results.

  In Section \ref{Intro}, the action (\ref{S1E3}) was introduced as
  a generalization of the actions (\ref{S1E1}) and (\ref{S1E2}).
  This generalized action is characterized by two functions $D$ and
  $H$, along with a matter Lagrangian described by a general `potential' $V$.
  The ambiguity in the definition of material sources relative to the
   potential $V$  can yield a variety of field equations, of which we
   explored two types, denoted I and II. The former set was obtained by
    taking the stress-energy  tensor density $T_{\mu\nu}$ to be the  variation
of the matter action
   with respect to the metric, whereas the latter incorporated the variation of
the matter action
    with respect to the dilaton field into its definition (as in (\ref{S1E7})).
   The actual type-II equations we employed incorporated  the constraint
(\ref{S1E6})
  $H(\Phi) = k\,D'(\Phi)$.  It is important to note that the restriction
(\ref{S1E6}) put on the
  type-II field equations is not the reason why
  the two sets of field equations have different dynamical properties.
  Even if type-I equations satisfy condition (\ref{S1E6}), they will
  not in general have the same dynamical properties as those induced
  by type-II equations. The two sets will have exactly the same
  dynamical properties only if condition (\ref{S1E6}) holds and
  the dilaton source vanishes because they have different interpretation
  of the sources. This confusion in interpretation arises as a
  consequence of the lack of knowledge of the potential density
  $V$ which appears in the action. As a result, the ambiguity can
  only be clarified once the potential density is explicitly specified.

  For both types of field equations, the weak-field approximation
  was calculated. It is interesting to notice that the roles of the
  dilaton and matter sources seem to be interchanged in the two sets
  of calculations. In the type-I calculations, the perturbation of
  the dilaton field depends only on the matter source and the trace
  of the metric perturbation depends upon both dilaton and matter
  sources. The reverse situation happens in the type-II calculation.

  The post-Newtonian expansion, as in General Relativity, can be
  carried out to an arbitrary high order for both types of field
  equations. It is quite clear in Section\ref{Stellar} that the
  stellar equilibrium equations for both types of field equations
  can be reduced into Newtonian equation in the Newtonian limit.

  If we compare the solutions of the dust-filled and radiation-filled
  cases, we will find that the solutions in the radiation-filled case
  are generally more complicated than those in the dust-filled case.
  This is the result we expected because there is no interaction
  between the particles, that is galaxies, in dust-filled model.
  During the course of developing a model for inflationary universe,
  even though the parametric functions $D$ can be found for each type
  of field equations, these functions are mathematically so
  complicated that the differential equations cannot be solved  analytically
when
  dust-filled or radiation-filled universes are considered.


\section*{Acknowledgements}
  This work was supported in part by the Natural Sciences
  and Engineering Research Council of Canada.

\end{document}